# Investigation of ultrashort laser excitation of aluminum and tungsten by reflectivity measurements

T. Genieys[1], M. Sentis[1], O. Utéza[1]

[1]*Aix-Marseille University, CNRS, LP3 UMR 7341, F-13288 Marseille, France*

Corresponding author: genieys@lp3.univ-mrs.fr

**Abstract**

We determine the laser-induced ablation threshold fluence in air of aluminum and tungsten excited by single near-infrared laser pulses with duration ranging from 15 fs to 100 fs. The ablation threshold fluence is shown constant for both metals, extending the corresponding scaling metrics to few-optical-cycle laser pulses. Meanwhile, the reflectivity is measured providing access to the deposited energy in the studied materials on a wide range of pulse durations and incident fluences below and above the ablation threshold. A simulation approach, based on the two-temperature model and the Drude-Lorentz model, is developed to describe the evolution of the transient thermodynamic and optical characteristics of the solids (lattice and electronic temperatures, reflectivity) following laser excitation. The confrontation between experimental results and simulations highlights the importance of considering a detailed description and evolution of the density of states in transition metals like tungsten.

## 1 Introduction

Femtosecond lasers have the ability to machine materials with good efficiency, minimized thermal budget and collateral effects [1, 2]. Nevertheless, the evaluation of observables to benchmark matter transformation is still scarce in the ultrashort regime (<< 100 fs) providing evident interest to this ongoing research. For instance, no determination of single-shot laser-induced ablation threshold fluence ($F_{th}$) can be found for most metals, including tungsten or aluminum studied here, at pulse durations as short as 15 fs. Measurements can be found at longer femtosecond pulse duration, most commonly in multi-pulse regime with the objective of surface structuration of metallic materials and enhancement of their surface properties for various scientific and industrial applications [3, 4].

Moreover, the development of predictive quantitative models for the description of the interaction is of great interest whether for a best control of material modification or to estimate the resistance of a given material to laser irradiation. The most widely employed is the two-temperature model (TTM) [5] which calculates the temporal and spatial evolution of the temperature of electron and ion subsystems following laser excitation. Improvement of modeling accuracy involves understanding complex phenomena at different time scales, from femtoseconds to nanoseconds. They include laser heating of the electrons and electron-electron energy exchange at the pulse time scale followed by progressive energy transfer to the lattice. In most materials including metals, the characteristic time for completing such energy transfer is in the range of a few ps to tens or hundreds of ps depending on the evolution of the electron collision frequency of the material following excitation [6-8]. Finally, when the energy deposition is intense enough, removal of material (ablation) takes place later on at the nanosecond time scale after release of the coupled laser energy into internal thermo dynamical energy (mechanical and thermal constraints) [7, 9, 10].

To provide experimental insights into laser heating of metals and further test the applicability of the TTM, the evaluation of the reflectivity of the excited sample during the interaction provides a useful observable. The description of reflectivity usually assumes that the pulse interacts with a free electronic population. In this case the Drude model can be applied. However, several works indicate that the electronic structure (Density of States, DOS) of metals plays a major role in the evolution of physical characteristics following electronic excitation [11-13]. In particular, in transition metals, the DOS is characterized by the presence of regions of high density of states around the Fermi energy, associated with d-band electrons, which can induce screening during the laser excitation. In those cases, the optical response of the material is not conveniently described by the Drude model and heating of the electronic population leads to non-trivial changes of electron-phonon coupling factor [11], electronic heat capacity [11] and optical properties [14].

To investigate the importance of the details of the excited material electronic structure, two metals are selected: aluminum (Al), a post-transition metal with a DOS distribution close to the free electron gas model [11] and tungsten (W), a transition metal, with a DOS distribution characterized by the presence of a large region of high density of states associated with the d-band above and below the Fermi energy [11]. The present manuscript is organized as follows. After the description of the experimental configuration in section 2, we introduce the theoretical background in section 3. Then, in section 4, the laser-induced ablation threshold fluence ($F_{th}$) of Al and W is experimentally determined for pulse durations ranging from 15 to 100 fs, providing ablation threshold metrics in a range of pulse duration poorly explored yet. Meanwhile, the evolution of reflectivity of Al and W samples integrated over the pulse is measured as a function of the incident laser fluence and for pulse durations ranging from 15 to 100 femtoseconds. In section 5, the experimental results are compared to simulation results obtained with the TTM and Drude or Drude-Lorentz models, providing evidence of the importance of taking into account the DOS when studying laser-metal interaction in the ultra-short regime.

## 2 Experimental arrangement and details

Experiments are performed in air at normal incidence in single-shot regime using the beam line 5a of ASUR platform at LP3 laboratory. This beam line delivers linearly polarized 30 fs pulses at 100 Hz repetition rate with a central wavelength of 800 nm ($\omega_l = 2.35 \times 10^{15}\ s^{-1}$, $\Delta\lambda \cong 760 - 840$ nm FWHM) and a maximum energy of 1 mJ with 1% rms fluctuations. Two experimental configurations were used, a detailed description being provided in [16-18]. In short, the first test bench allows a variation of the pulse duration in a range between 30 and 100 fs (FWHM) by pre-chirping the beam through compressor grating adjustments. The second setup provides access to shorter pulse durations using two BaF$_2$ crystals set in a vacuum tube and cross-polarized wave generation to broaden the spectrum (720 – 880 nm), and compression based on chirped mirrors and a pair of fused silica wedges [16, 19]. Pulses on target have a maximum energy of ~ 45 µJ with duration down to 15 fs (FWHM) and 2.5% rms energy fluctuations. In both setups, the incident energy is controlled by the combination of a half-wave plate and a set of four thin Brewster polarizers whose dispersion is accounted for using pre-chirping strategy. The measurement of the pulse duration is done before the off-axis focusing parabola by means of second-order autocorrelation. Finally the pulses are focused using an off-axis parabola of effective focal length of 50.8 mm (15 fs case) and of 152.4 mm (30 – 100 fs cases) respectively. The sizes of the incident beam on the parabolic mirror are slightly adjusted by means of an iris diaphragm (losses < 10%) to get similar f-number and therefore similar beam dimensions in the focal plane in which the target is thereafter positioned. The spatio-temporal profile of the laser beam is Gaussian, characterized by a radius $\omega_0$=11 µm (for 30 – 100 fs) and $\omega_0$=10.2 µm (for 15 fs) (at $1/e^2$ in the focal plane).

Aluminum (GoodFellow AL0065, high purity 99.99%, thickness of 0.5 mm) and tungsten (GoodFellow W000375, high purity 99.95%, thickness of 2.0 mm) are used as test materials. Thermo-physical properties of these materials relevant to this study are summarized in Table 1.

| Material | **Aluminum** | **Tungsten** | **References and Remarks** |
|---|---|---|---|
| Electronic configuration | [Ne]$3s^2 3p^1$ | [Xe]$4f^{14} 5d^4 6s^2$ | [20] |
| Mass of atom (a.m.u.) | 26.98 | 183.84 | [20] |
| Atomic density ($n_{at}$, m$^{-3}$) | 6.02×10$^{28}$ | 6.32×10$^{28}$ | [20] |
| Number of free electrons/atom | 3 | 2 | [21]. These numbers are kept constant as a function of the electronic temperature. |
| Fermi energy ($E_F$, eV) | 10.8 | 9.2 | $E_F = \frac{m_e v_F^2}{2}$, with the Fermi velocity $v_F = \hbar k_F / m_e$ and the Fermi wave vector $k_F = (3\pi^2 n_e)^{1/3}$. |
| Plasma frequency $\omega_p (s^{-1})$ | 2.28×10$^{16}$ | 2.01×10$^{16}$ | $\omega_p = \sqrt{\frac{e^2 \, n_e}{m_e \, \varepsilon_0}}$ |
| Density (g.cm$^{-3}$) | 2.70 | 19.3 | [20] |
| Debye temperature (K) | 426 | 405 | [22] |
| Melting temperature (STP) (K) | 933 | 3695 | [20] |
| Boiling temperature (STP) (K) | 2792 | 5828 | [20] |
| Enthalpy of fusion and evaporation (kJ.mol$^{-1}$) | 10.79-294.0 | 35.4-824.0 | [20] |
| Volumetric heat capacity at 300 K ($C_p$, J.m$^{-3}$K$^{-1}$) | 2.46 | 2.51 | [20] |
| Electronic thermal conductivity, for 300 K ($\kappa_e$, W.m$^{-1}$.K$^{-1}$) | 237 | 174 | [20] |

**Table 1** Electronic and thermo-physical properties of aluminum and tungsten.

Beforehand, the reflectivity of both samples is measured using the 800 nm (1.55 eV) – 30 fs collimated beam at very low incident fluence for which no significant changes in temperature (electronic and lattice) and permittivity (thus conductivity) occur. Those irradiation conditions define the initial optical properties of the material further denoted "unperturbed material". The optical response of the material is given by Fresnel formulas and the complex permittivity of the material. For tungsten, a reflectivity of $R_{0,coll}$ = 0.497 +/- 0.012 is measured, which is in very good agreement with calculations using its dielectric function coefficients at 800 nm ($R_{0,calc}$ = 0.501 [23]). For aluminum, a reflectivity of $R_{0,coll}$ = 0.773 +/- 0.012 is measured, which is lower than the reflectivity obtained in [22] $R_{0,calc}$ = 0.871. The main reason for this difference is the scattering losses through diffuse reflectivity. Indeed, in this collimated geometry of measurement with a low collection solid angle, only the specular reflected energy by the sample is monitored. The diffuse reflectivity losses can be estimated with the formula expressing the reflectivity scattered $R_{scatter}$ by a rough surface (with Ra characteristics) to the reflectivity of a perfectly smooth surface $R_{smooth}$ [24]:

$$R_{scatter} / R_{smooth} \approx \exp\left[-\left(\frac{4\pi Ra}{\lambda}\right)^2\right] \quad (1)$$

Both samples were polished before the experiments and a roughness $Ra$ = 20 nm (Al) and $Ra$ = 8 nm (W) was measured by atomic force microscopy (PSIA XE-100). Using eq. 1 for aluminum, a value of $R_{smooth-Al}$ = 0.862 is obtained, which is in good agreement with [23]. In case of tungsten, a value of $R_{smooth-W}$ = 0.505 is obtained using eq. 1, which is also in very good agreement with [23]. For both metals, in the following calculations we will use the values of $R_{smooth}$ (labelled $R_0$ in the following).

In order to study the evolution of reflectivity, the incident and reflected energy of the pulse are measured by photodiodes for each single shot at a given energy (fluence). The reflected signal is collimated by the off-axis parabola and redirected by two transport dielectric mirrors whose spectral bandwidths correspond to the 15 fs and 30-100 fs spectral content of the ASUR beam lines to ensure spectrally isolated measurements of the reflected pulse. The experimental arrangement, which allows collecting a high solid angle, captures the entire reflected signal, which was verified by replacing the photodiode by a beam analyzer CCD camera (detector surface: 6.5 mm x 6.5 mm). Moreover, a short focal lens is positioned in front of the photodiode to provide robustness of the measurement to small misalignment. This experimental arrangement allows the determination of the reflectivity $R$ integrated over the pulse duration. The evolution of reflectivity is further normalized by setting $R = R_0$ at very low incident fluence corresponding to the initial unperturbed state of the material as defined before.

# 3 Theoretical background and modeling

## 3.1 Drude-Lorentz equation

As a first approximation, we use the Drude-Lorentz model to calculate the dielectric function of the studied metals to further access to the reflectivity. The Drude-Lorentz model provides a description of the interaction including bound and free electron optical response. It accounts for the optical properties of the material in many situations of irradiation, especially when the material can be considered unperturbed or slightly excited (F << $F_{th}$). In the framework of the Drude-Lorentz model, the dielectric permittivity is equal to [23]:

$$\varepsilon = \varepsilon_{Drude} + \varepsilon_{Lorentz} = [1 - \frac{f_0 \omega_p^2}{\omega(\omega - i\nu)}]_D + [\sum_{j=1}^{k} \frac{f_j \omega_p^2}{(\omega_j^2 - \omega^2) + i\omega\Gamma_j}]_L \quad (2)$$

In this approach, electrons exposed to an external electric field are treated as oscillating particles with a first term describing the contribution of delocalized free electrons (Drude model) and a second term accounting for the response of bound electrons (Lorentz model). Considering the first term, $\omega_p$ indicates the plasma frequency associated with s/p intra-band transitions, $f_0$ is the oscillator strength, $\nu$ the collision rate and $\omega$ the laser frequency. The second term is described by the Lorentz component where $f_j$, $\omega_p$, $\omega_j$, and $\Gamma_j$ are the oscillator strength, the plasma frequency, the frequency and the scattering rate of the harmonically bound electrons excited via the inter-band transition j, respectively. All the parameters entering in the calculations for aluminum and tungsten at room temperature are listed in [23]. Afterwards, the complex index coefficients ($n + ik$) and the reflectivity are calculated using the following formulas: $R = \left|\frac{\sqrt{\varepsilon}-1}{\sqrt{\varepsilon}+1}\right|^2$, and: $n = \sqrt{\frac{\varepsilon_r + \sqrt{\varepsilon_r^2 + \varepsilon_i^2}}{2}}$, $k = \sqrt{\frac{-\varepsilon_r + \sqrt{\varepsilon_r^2 + \varepsilon_i^2}}{2}}$, in which $n$ and $\varepsilon_r$ and $k$ and $\varepsilon_i$ are respectively the real and imaginary parts of the complex refraction index and of the permittivity. The Drude model (contained in the Drude-Lorentz model as shown in equation 2) does not take into account the full electronic structure of metals. It only considers the laser absorption on the delocalized free electron population and it gives a useful and reasonable approximation to describe the optical response of several metals. In particular, the Drude model is applicable for aluminum whose DOS distribution corresponds to the free electron gas model and thus presents a high number of delocalized vacant states above the Fermi energy accessible to laser coupling (see also section 4.2). In the following calculations, the Drude model approximation is used for aluminum.

To benchmark the initial optical response of both materials, we first consider the unperturbed reflectivity. For the unexcited cold solid at room temperature (300 K), the collision rate is governed by the electron-phonon collision rate [25]. In the range from the Debye temperature up to the melting point, the electron-phonon collision rate is approximated by the following expression [26]:

$$\nu_{e-ph} = \frac{3}{2} C_\omega \frac{(k_B T_i)}{\hbar}, \qquad (3)$$

where $k_B$ is the Boltzmann constant, $T_i$ the lattice temperature, $\hbar$ the reduced Planck constant, and $C_\omega$ a dimensionless proportionality coefficient suitably determined from dedicated experiments. Afterwards, using the unperturbed reflectivity measured previously ($R_{0,Al} \cong 0.862$ and $R_{0,W} \cong 0.505$) and the Drude model for aluminum, and the Drude-Lorentz model for tungsten, we determine $\nu_{e-ph} \cong 1.31\times10^{15}$ s$^{-1}$ for Al and $\nu_{e-ph} \cong 1.91\times10^{14}$ s$^{-1}$ for W. It also corresponds to the complex refractive index: $n_{Al} = 3.02 + i8.48$ and $n_W = 3.63 + i2.81$ which are in good agreement with tabulated values in [20, 27] in the corresponding spectral range ($n_{Al,1.6eV-775nm} = 2.62 + i8.59$ and $n_{W,1.6eV-775nm} = 3.67 + i2.68$ [27]).

These values of electronic collision frequency determined for the unexcited cold solid will be used as a starting input parameter to calculate the evolution of reflectivity following absorption of an ultrashort laser pulse. When the electronic temperature increases, the collision frequency is dominated by electron-electron collisions. Fermi's theory of liquids allows describing the evolution of an excited electron in an electron population thermalized at the temperature $T_e$ [6, 28]:

$$\nu_{ee} = \frac{\pi^4 k_B^2 \sqrt{3}}{256} \frac{\omega_p}{E_F^2} T_e^2, \qquad (4)$$

where $E_F$ is the Fermi energy (see table 1). To fully describe the energy exchanges occurring during interaction (within the cold solid limit $T_e < T_F$, with $T_F = E_F/k_B$, the Fermi temperature), the effective collision frequency $\nu_{eff}$ is considered being equal to the sum of the electron-electron and electron-phonon collision frequencies.

### 3.2 Two-Temperature Model

The description of the energy exchanges between the electronic population of temperature $T_e$ and the lattice temperature of temperature $T_i$ is obtained in the frame of a 1D-space (x, being the coordinate along the optical axis, $x = 0$ indicates the surface of the material) and time dependent approach with two coupled equations [5] (for sake of simplicity, implicit notation, e.g. $T_e$, instead of $T_e(x,t)$, is used):

$$C_e \frac{\partial T_e}{\partial t} = \frac{\partial}{\partial x}\left(\kappa_e \frac{\partial T_e}{\partial x}\right) - g(T_e - T_i) + S \qquad (5)$$
$$C_i \frac{\partial T_i}{\partial t} = g(T_e - T_i)$$

$C_e$ and $C_i$ stand for the heat capacities of the electrons and the lattice, $g$ is the electron-phonon coupling factor and $\kappa_e$ is the electronic thermal conductivity. Since the energy transport is mainly provided by electrons in metals, the term concerning the thermal conductivity of the lattice is omitted in equation 5. Moreover, every physical quantity in eqs. 5, except $C_i$, is a function varying in space and time, as specified below. The electronic thermal conductivity varies according to the electronic collision frequency $v_{eff}$: $\kappa_e = \frac{V_F^2 C_e}{3v_{eff}}$ [11], with $V_F$ the Fermi velocity. $S$ is the source term referring to the amount of laser energy absorbed: $S(x,t) = \alpha(1-R)I(t)\exp(-x\alpha)$, where $\alpha$ is the inverse of the optical penetration depth ($l_{opd} = \frac{c}{2\omega_l k}$), $R$ the reflectivity, and $I(t)$ the laser intensity for which Gaussian temporal shape is considered. As a first approximation, the absorption coefficient $\alpha$ and the reflectivity are taken constant. This assumption is true for fluences below and equal to the ablation threshold fluence because the reflectivity does not vary significantly ($R \cong R_0$ for $F \leq F_{th}$) for all the pulse durations as it experimentally demonstrated later (see figure 2).

In the TTM model [5], thermalization of the electronic population into a Fermi-Dirac distribution allowing the definition of an electronic temperature ($T_e$) is assumed to be instantaneous. During the interaction with a femtosecond laser, the electronic population can be easily excited which causes a shift of the Fermi level and a modification of electronic heat capacity and electron-phonon coupling factor. The evolution of these characteristics can differ greatly from the free electron gas model (especially for W) and has been calculated in [11] as a function of the electronic temperature. For our calculations, $C_e$ and $g$ are taken from [11] using a fitting procedure based on polynomial functions. Finally, as the main purpose of this calculation is to evaluate the evolution of the material characteristics during the pulse, no plasma expansion is taken into account.

## 4 Results

### 4.1 Laser-induced ablation threshold fluence in ultrashort regime

The laser-induced ablation threshold fluence ($F_{th}$) of aluminum and tungsten is determined in single shot regime for pulse duration of 15, 30, 50 and 100 fs. We define ablation as removal of material. To infer the ablation threshold energy, we use the classical diameter-regression technique [29], considering at different irradiation levels the evaluation of the interaction of the laser with the metal sample (ablated diameter measured by confocal microscopy) and the Gaussian distribution of the pulse (see figure 1). The different slopes between 15 fs and the other pulse durations are due to different focusing geometries used in the two test-benches.

The ablation threshold fluence values obtained for every pulse duration are shown in table 2. The accuracy of the waist measurement on the CCD camera is evaluated at +/- 0.25 μm and the shot-to-shot energy fluctuation is 2.5% at 15 fs and 1% for the other pulse durations. The measurement uncertainties for the fluence threshold values are therefore +/- 0.023 J/cm² at 15 fs and +/- 0.018 J/cm² for the other pulse durations. Therefore, taking into account the uncertainty of the measurements, the laser-induced ablation threshold fluence of Al and W does not vary for pulse duration changing from 15 to 100 fs. It is equal to $F_{th,Al} \cong 0.235$ J/cm² for aluminum and $F_{th,W} \cong 0.53$ J/cm² for tungsten. For aluminum, a fluence of 0.12 J/cm² was measured in [30] with similar laser parameters (800 nm, 100 fs). The difference with the value measured in table 2 may originate from a different approach to estimate the threshold. In [30], the value is deduced from multi-shot measurements which can lead to uncertainties about the threshold value due to incubation effects. In [31], it was measured a much higher ablation threshold fluence in single-shot regime (800 nm, 100 fs): $F_{th} \cong 0.9$ J/cm². However, this experiment was performed in vacuum, so in experimental conditions which are not directly comparable to our experiment done in air. For instance, two times higher ablation threshold fluences were measured in vacuum for identical samples in ps regime (12 ps, 532 nm) when comparing thresholds in vacuum and in air [32].

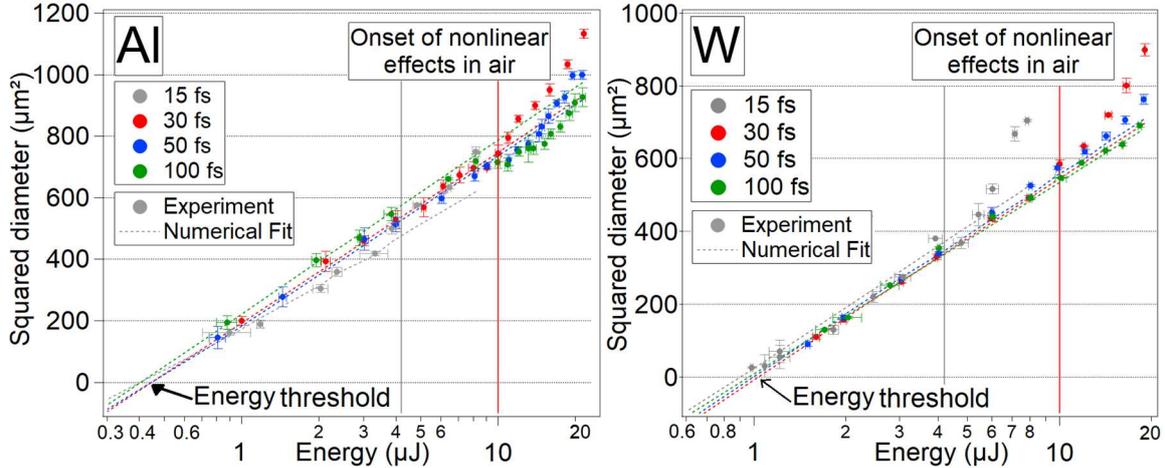

**Fig. 1** Evolution of squared diameter D² versus incident laser energy for every pulse duration. Each data is averaged over 8 measurements, the plotted vertical error bars being the standard deviation and the horizontal error bars the shot-to-shot energy fluctuations measured on the incident photodiode. The grey and red vertical solid lines correspond to the onset of nonlinear effects in air at 15 and 30 fs as it was measured on the two setups [16-18]. The ablation threshold energy is determined when the fit intercepts the criterion D² = 0. The points at high incident energy are not included in the fit because of the progressive loss of control of the incident intensity on target in presence of nonlinear effects and risks of biased interpretation.

| Pulse Duration | 15 fs | 30 fs | 50 fs | 100 fs |
|---|---|---|---|---|
| Aluminum, $F_{th}$ (J/cm²) | 0.23 | 0.24 | 0.240 | 0.23 |
| Tungsten, $F_{th}$ (J/cm²) | 0.52 | 0.54 | 0.530 | 0.53 |

**Table 2** Laser-induced ablation threshold fluence of aluminum and tungsten for 15, 30, 50 and 100 fs. The ablation threshold energy $E_{th}$ is deduced from the diameter regression technique (see figure 1). To calculate the corresponding peak fluence ($F_{th} = \frac{2E_{th}}{\pi w_0^2}$) we use $\omega_0$ the radius of the focal spot at 1/e² measured by imagery (see paragraph 2): $\omega_0$=11 μm (for 30 – 100 fs) and $\omega_0$=10.2 μm (for 15 fs).

The quantitative value measured for tungsten (0.53 J/cm²) is in good agreement with single-shot results from previous reports where similar irradiation conditions were used (100 fs, 800 nm, $F_{th}$ = 0.55 ± 0.1 J/cm², [33]). However, in [33], the experiments are performed behind the focal point to get a large spot size making less straightforward the precise knowledge of the beam spot size at the plane of the test and also inducing a higher sensitivity to nonlinear effects.

Finally, as an important outcome of this work, we conclude that for both metals the ablation threshold was determined constant over the tested pulse duration range allowing us to extend the ablation threshold scaling metrics observed in sub-picosecond regime (≤ 100 fs – 1 ps) to few–optical-cycle laser pulses (till ≅ 15 fs).

**4.2 Measurement of reflectivity in ultrashort regime**

The evolution of the reflected energy fraction of the pulse as a function of the incident fluence is shown in figure 2 for pulse durations of 15, 30, 50 and 100 fs.

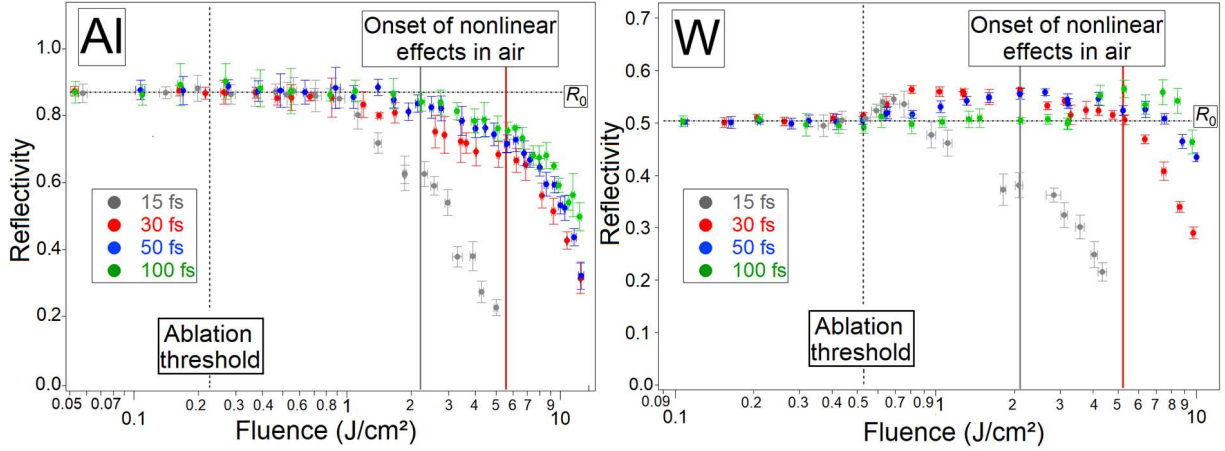

**Fig. 2** Evolution of reflectivity of the Al and W samples integrated over the pulse as a function of the incident fluence for pulse durations from 15 to 100 fs. The horizontal line corresponds to the unperturbed reflectivity $R_0$. The vertical dot line corresponds to the ablation threshold and the other grey and red vertical solid lines have the same meaning as in figure 1.

In both experiments, for incident fluences equal or lower than the ablation threshold, the reflectivity remains unchanged with respect to the unperturbed reflectivity $R_0$. When the incident fluence increases above the threshold, a decrease in reflectivity is measured for Al (figure 2, left). It is initiated at higher fluence for longer pulse duration. On the other hand, for the tungsten sample, when the fluence increases above the threshold, an increase in reflectivity above $R_0$ is first measured. It is interesting to note that the increase of tungsten reflectivity for moderate excitations has been calculated in [14] using *ab initio* molecular dynamic simulations. Afterwards, at high fluences, a large decrease (similar to what was measured for Al) is also detected. Note that the augmentation of reflectivity seen for tungsten is not observed at the same fluence for the four pulse durations tested. Indeed, the increase occurs at higher incident fluence when the pulse is temporally stretched (see figure 2, right).

To explain the observed behavior, at first we consider the influence of the evolution of the collision frequency on the reflectivity. Upon laser excitation, the effective electron collision frequency swiftly increases according to the increase of the electronic temperature raising in turn the electron-electron collision frequency (see eq. 4) [8, 25]. As energy transfer to the lattice and related lattice temperature elevation are completed much after the termination of the pulse (see thereafter figures 5 and 8), the electron-phonon collision frequency keeps almost unchanged at the pulse time scale. Then it is interesting to plot the evolution of reflectivity as a function of collision frequency for aluminum and tungsten in the framework of the Drude and Drude-Lorentz models for the two metals (see figure 3). For that, we used eqs. 2-4 and the coefficients in [23]. In this calculation, we assume that the coefficients obtained in [23], describing the response of the bound electrons, does not significantly vary during the time of our measurement corresponding to the reflectivity integrated over the duration of the pulse. To support this hypothesis, we again argue that only weak perturbation of the material takes place at the pulse time scale. Indeed, as shown later in section 5, the increase of the lattice temperature only occurs after the termination of the pulses and, for fluences around the ablation threshold, the electronic temperature remains lower than the Fermi temperature (cold solid regime).

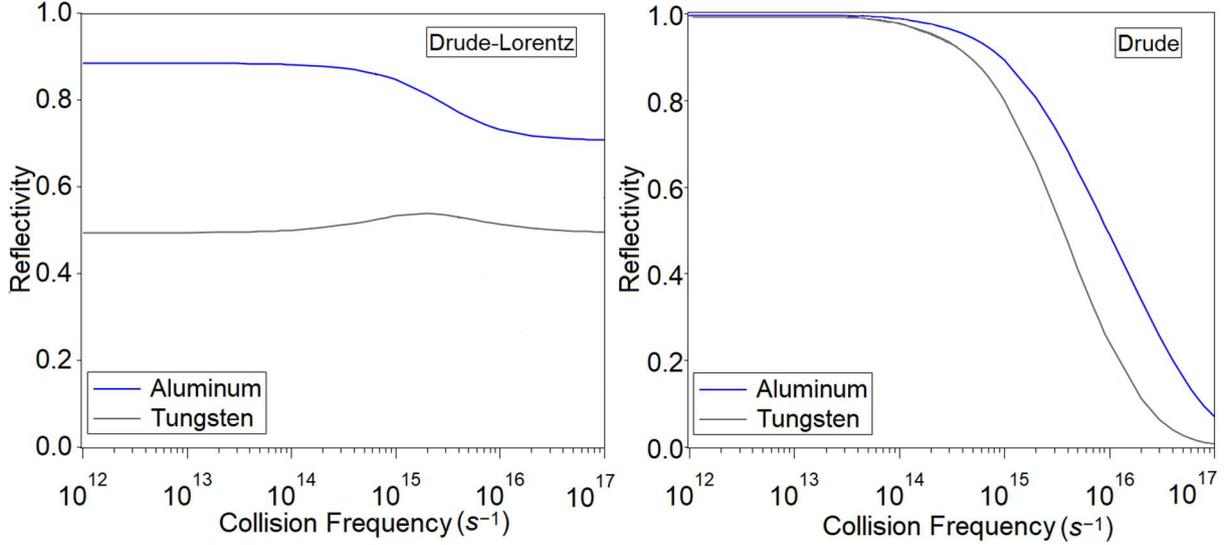

**Fig. 3** Evolution of reflectivity as a function of the electronic collision frequency for Al and W, calculated with Drude-Lorentz (left) and Drude models (right). Calculations performed using eq. 2 and parameters listed in [23].

It immediately appears from figure 3 that the increase of reflectivity observed above the ablation threshold in case of tungsten (see figure 2) can only be explained in the framework of the Drude-Lorentz model, incorporating the contribution of the bound (d-band) electrons to describe the optical response. Moreover, as the increase of the effective electron collision frequency is sensitive to the electronic temperature, depending in turn to the intensity [34], it is expected to observe changes of the reflectivity at lower fluences when the pulse duration is shortened.

Afterwards, for higher fluences, a decrease in reflectivity is measured on both samples (see figure 2) which is reproduced by the two modeling approaches (figure 3). However, the variation predicted by the Drude-Lorentz model is less important than in case of the Drude model. For incident fluences well above the ablation threshold, large decreases of reflectivity (> 10 %) are measured for both metals (see figure 2). In those conditions (for fluences many times above the ablation threshold), the optical response is dominated by the delocalized s/p-band electron population for both studied metals. As a result, the evolution of the reflectivity can be described by the Drude model (figure 3, right), approximating a free electron gas. In the framework of this model, the decrease in reflectivity is related to an increase of the electronic collision frequency, which itself varies according to the electronic temperature following pulse absorption.

Moreover, at a given fluence, for shorter pulse durations, the intensity is higher, and the linear absorption (Inverse Bremsstrahlung) is proportional to the intensity of the pulse. Electronic heating is therefore more efficient for shorter pulse duration. The electronic temperature increases earlier during the pulse and the reflectivity starts to decrease earlier as well. When the trailing edge of the incident energy interacts with the sample surface, a larger amount is absorbed, increasing the electronic temperature even more. The earlier the electronic temperature required for a drop in reflectivity is reached during the pulse, the more the measured time-integrated reflectivity will change, which explains why larger reflectivity changes are measured at short pulse durations (fig. 2).

## 5 Discussion: importance of DOS electronic structure

After having given a general description of laser coupling in a wide range of excitation for both metals, we now concentrate to the analysis of the energy deposition process for the two metals, highlighting the importance of their electronic structure to laser beam coupling in ultrashort regime.

### 5.1 Aluminum

The evolution of the density of states of aluminum as a function of electronic energy (Fig. 4) is close to the free-electron gas model, evolving with a dependence of $\sqrt{E}$ ($E$ being the energy of an electron). For this metal, delocalized vacant states above the Fermi energy are immediately available and accessible to laser coupling and free-electron intra-band collisions are dominant. The Drude model can therefore be applied for the calculation of reflectivity.

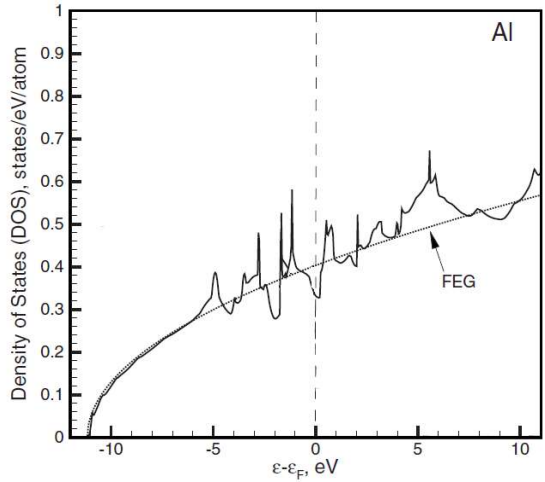

**Fig. 4** Density of states of aluminum. Taken from [11]. FEG acronym refers to the Free-Electron Gas distribution model.

In aluminum, the time for thermalization of the electronic population was estimated at ~2 fs with a pulse of 10 fs at 800 nm and an absorbed fluence F = 0.7 mJ/cm² [6]. This fluence is much lower than the smallest fluence conditions explored in our experiment ($F_{inc}$ = 50 mJ/cm² with corresponding absorbed fluence $F_{abs} = (1 - R_0) F_{inc}$ = 7 mJ/cm²). In [6], it was also shown that the thermalization time decreases when the absorbed fluence increases. For all our irradiation conditions, it is therefore much less than 2 fs and thermalization of the electron subsystem can be considered instantaneous. In that case, the two-temperature model can be directly applied to describe the interaction and to calculate the evolution of the electronic ($T_e$) and lattice ($T_i$) temperatures.

This calculation is performed at the material surface for an incident fluence of $F_{th,Al}$ = 0.235 J/cm² corresponding to the measured ablation threshold. The reflectivity is taken equal to the unperturbed reflectivity $R_0$ for estimating the source term. This is consistent with small variations of the transient reflectivity keeping the integrated value unchanged as it was measured at the ablation threshold. Within this assumption, the exact correspondence between the energy absorbed measured experimentally ($E_{abs}$ = 1 - $E_{Reflected}$) and the total laser energy considered as the source term in the TTM model is also preserved. During this calculation, the electronic heat capacity $C_e$ and the coupling factor $g$ are taken from [11] using a fitting procedure. The thermal conductivity of the electrons $\kappa_e$ is calculated at each step with the formula: $\kappa_e = \frac{V_F^2 C_e}{3\nu_{eff}}$, $V_F$ being the Fermi velocity and $\nu_{eff}$ the effective collision frequency. Results of the TTM for pulse durations ranging from 15 to 100 fs are shown in figure 5.

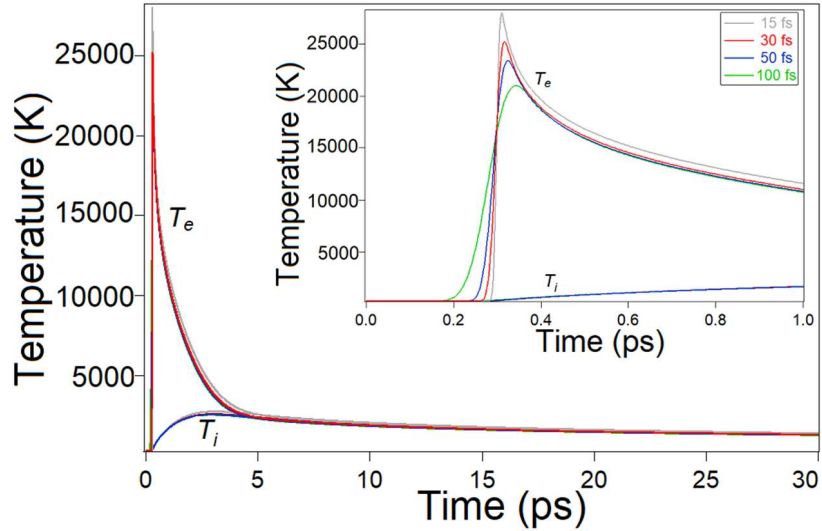

**Fig. 5** Temporal evolution of electronic ($T_e$) and lattice ($T_i$) temperatures for a laser fluence equal to the ablation threshold ($F_{th,Al}$ = 0.235 J/cm²) and pulse durations ranging from 15 to 100 fs for aluminum (Results of TTM calculations, using eqs. 5). The pulses are centered at 0.3 ps. The inserted frame corresponds to the evolution during the first picosecond.

During the pulse a sudden increase in the electronic temperature is observed. The maximum value reached is slightly higher for the shortest pulse durations (i.e. for the highest intensities) with a temperature $T_e$ ~ 26000 K (2.24 eV) for a pulse of 15 fs and $T_e$ ~ 20000 K (1.72 eV) for a pulse of 100 fs.

The maximum temperature reached by the lattice is the same for the different pulse durations. Assuming $R$ constant in the TTM, while varying $C_e$, $g$, and $\nu_{eff}$ as a function of the electronic temperature as determined in previous sections 3.1 and 3.2, we can deduce that changing the pulse duration has an influence on the dynamics of the energy transfer within the electron subsystem and to the lattice. However, the changes of the electronic temperature depending on pulse duration appear to be limited both in terms of amplitude and in time. As shown in figure 5, the electronic temperature is nearly identical at the end of the laser pulse for all pulse durations studied. Electron transport, and energy exchange and transfer within the electron subsystem and to the lattice thus differ only marginally according to the pulse duration. As a consequence, despite reaching different maximum values of $T_e$ and recalling that the same total amount of absorbed energy is measured whatever the pulse duration, the maximum temperature reached by the lattice is nearly identical for all four pulse durations tested.

The maximum lattice temperature value reached is $T_i \sim 2800$ K, which is higher than the melting temperature and close to the boiling temperature (see table 1). Among the mechanisms of material ablation by a femtosecond pulse, phase explosion has been recognized as dominant and most probable, especially at high incident fluences [9, 35-37]. Phase explosion implies overheating of the surface focal volume of the target to high lattice temperature beyond the limit of thermodynamic stability of the target material [9, 35, 38], further followed by explosive ejection of the heated surface volume into a two-phase mixture of liquid droplets and vapor. In a quantitative perspective, it was determined that phase explosion develops as the lattice temperature reaches ~ 90% of the critical temperature $T_c$ [9, 35, 38]. For Aluminum, critical temperature values ranging from 5100 K to 8800 K can be found in the literature [39]. The lattice temperature determined using the TTM model is lower than this value which suggests that phase explosion is not the main mechanism yielding to ablation of Al in ultrashort regime. More probably, ablation at the threshold fluence ($F_{th,Al}$ = 0.235 J/cm²) would correspond to photomechanical ablation, typically occurring for incident fluences lower than phase explosion [31, 38]. Photomechanical ablation, or spallation, supposes the formation of tensile stress with sufficient strength to induce a mechanical rupture or decomposition of the laser heated surface volume of the material. Even if such ablation mechanism is favored when using ultrashort laser pulses, detailed calculations would be required to verify if the condition of stress confinement would be reached in our irradiation conditions [38].

To extend the analysis, the evolution of reflectivity has been calculated (fig. 6) using the Drude model and the values of $T_e$ and $T_i$ obtained with the TTM (fig. 5). To do this, we assumed that the absorption coefficient is constant which appears to be reasonable because the changes in reflectivity during the pulse are small (see figure 6) and experimentally, they are even not detected. Moreover, the evolution of the effective electronic collision frequency over time ($\nu_{eff} = \nu_{eph} + \nu_{ee}$) is calculated using formula of the electron-phonon and electron-electron collision frequency (eqs. 3 and 4).

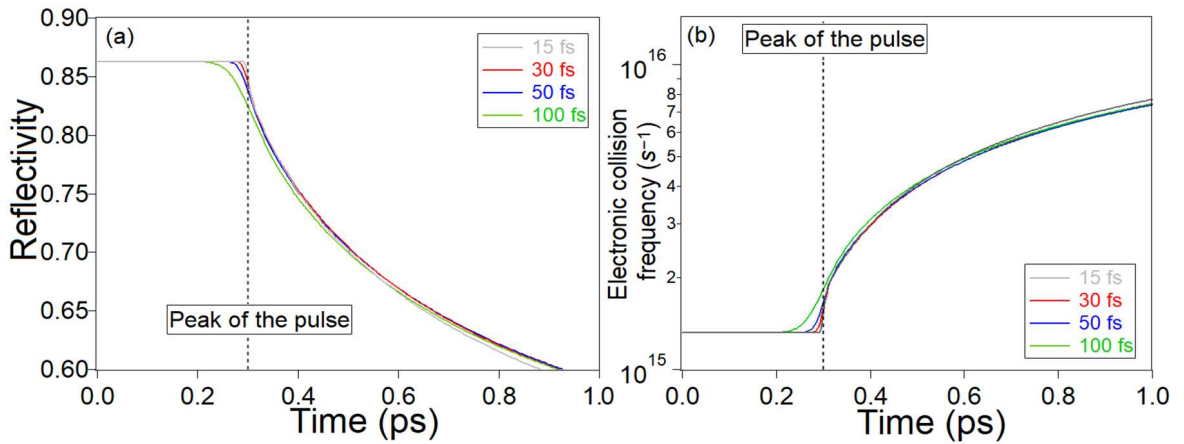

**Fig. 6** (a) Time evolution of the reflectivity of aluminum irradiated at the ablation threshold fluence ($F_{th,Al}$ = 0.235 J/cm²). (b) Corresponding time evolution of the electronic collision frequency. The pulses are centered at 0.3 ps. Calculations are done with the Drude model, based on the values of $T_e$ and $T_i$ obtained with the TTM.

The drop of reflectivity is the consequence of the increase of the effective collision frequency (see figure 6). At first, during the laser pulse, it is the result of the swift augmentation of the electron-electron collision frequency due to heating of the electronic population. On a longer time scale the electron population cools down; however the lattice temperature rises due to progressive deposited energy transfer to the lattice (fig. 5) yielding to an increase of the electron-phonon collision frequency and therefore contributing to sustain the effective collision frequency to high values.

In our experiment, the reflectivity measurement gives access to the integral of the reflectivity during the pulse duration. To compare our simulation with the experimental results, the integral of the reflectivity as a function of time must be calculated. From the value of $R$ plotted against time in figure 6, we calculate the value of the reflected power during the pulse, assuming a constant spatial distribution over the beam diameter. Then, the integral of power over the pulse duration gives the value of the reflected energy, allowing us to have access to the integrated value of reflectivity which can be compared to the measured observable. This calculation was performed for the four pulse durations at the ablation threshold fluence (0.235 J/cm²), resulting in an integrated reflectivity value $R \sim 0.85$ in each case. This is in good agreement with the experimental measurements ($R_0 \cong 0.862 \pm 0.012$), taking into account the uncertainty of this measurement.

**5.2 Tungsten**

The DOS distribution of W (figure 7) is more complex than the one of Al. The free electron gas approximation and Drude model cannot be applied as already remarked in section 4.2. Indeed, on figure 7, the presence of a region of high density of states associated with the d-band located from -6 eV to +4 eV around the Fermi level can be observed. However, the Fermi level is located in a local dip of the electron DOS distribution. The densely populated levels (on either side of the Fermi level) are separated by an energy, which can be compared to a pseudo band-gap ranging from $E - E_F \sim -1.20$ eV to $E - E_F \sim 0.75$ eV [11, 13, 14].

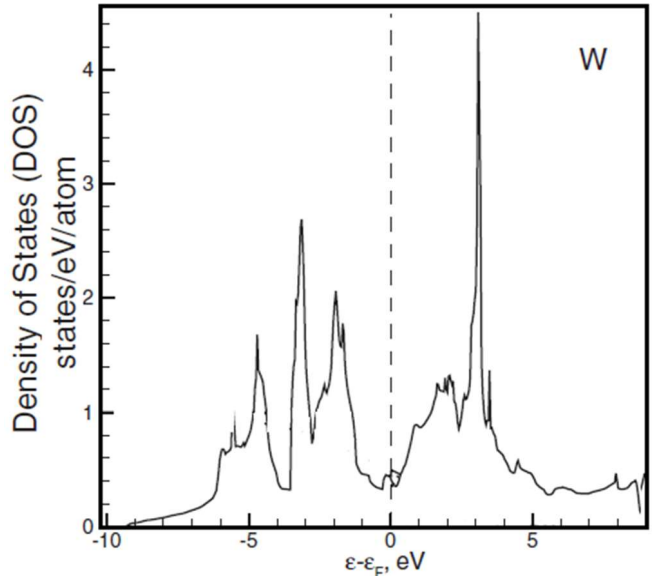

**Fig. 7** Density of states of tungsten. Taken from [11].

Upon laser excitation, the heating of the electron DOS population occurs in two possible ways. Considering the DOS distribution of tungsten initially populated up to the Fermi energy, a first absorption channel for photon coupling is provided by electron populated states far from the Fermi level ($E - E_F < -1.20$). These intra d-band transitions are not very effective because the electrons are excited to energy levels below 0.75 eV where only a few states are available. To overcome the pseudo band-gap and reach vacant levels where the density of states is higher, these electrons have to absorb at least two photons. The transitions from electron DOS states close to the Fermi level ($E - E_F >$ -1.20) are more probable because electrons can be promoted to vacant states available in numbers, beyond the pseudo band-gap. However, the DOS in this energy range (-1.20 eV $< E - E_F <$ 0 eV) is low (< 0.4 states/eV/atom) and these transitions are limited. At the ablation threshold fluence, the sample of tungsten absorbs an energy of ~ 0.5 µJ, which corresponds to the absorption of ~ $2\times10^{12}$ photons at 800 nm. We take here the data obtained on the setup used for experiments at 30, 50 and 100 fs; however same conclusions would apply to the 15 fs case. So, assuming that the absorption takes place in a cylindrical volume of depth $l_{opd}$ = 22.7 nm (using the imaginary part of the complex refraction index determined in section 3.1) and of radius $\omega_0$ = 11 µm, this corresponds to a photon density of $2.2\times10^{23}$ cm$^{-3}$. The free electron density (on which laser absorption occurs) is initially $n_e$ = $1.26\times10^{23}$ cm$^{-3}$. Therefore, during the interaction with a pulse at the ablation threshold, electrons absorb sequentially an average of 1.7 photons by Inverse Bremsstrahlung, which is insufficient to overcome the pseudo band-gap by effectively populating the high density of states available at higher energies. Therefore, during the absorption of a pulse at low or moderate excitation (F $\cong$ F$_{th}$) a weakly disturbed electronic distribution resulting from d-band screening is assumed. In these excitations conditions, the optical response remains dominated by the contribution of the bound d-band electrons in agreement with the experimental results ($R_{meas} \cong R_0$). In [15], it is shown that the excitation of a transition metal by an ultra-short pulse leads to the creation of two subsystems of electrons (associated respectively with the d and s/p bands) thermalized to two different temperatures. The equilibrium of the whole electronic system is reached after the end of the pulse, making the TTM inadequate in that case. This is exactly the situation encountered for nickel upon femtosecond irradiation [15, 40]. In the specific case of tungsten, the optical response is in fact dominated by the state occupation changes related to d-band electrons at low or moderate excitation (F $\cong$ F$_{th}$). We thus assume a simplified approach for TTM calculations in this excitation range, neglecting the contribution of the s/p bands electrons. So we do not seek to estimate the time of thermalization of the whole electronic population. Rather, we assume an instantaneous thermalization of the latter which appears to well reproduce the experimental results as shown later.

    The calculation is performed at the material surface for an incident fluence of F$_{th,W}$ = 0.53 J/cm² (ablation threshold fluence) with a reflectivity equal to $R_0$ = 0.505. As before, the electronic heat capacity $C_e$ and the coupling factor $g$ of tungsten are taken from [11] using a fitting procedure. The results are shown on figure 8 for the four pulse durations.

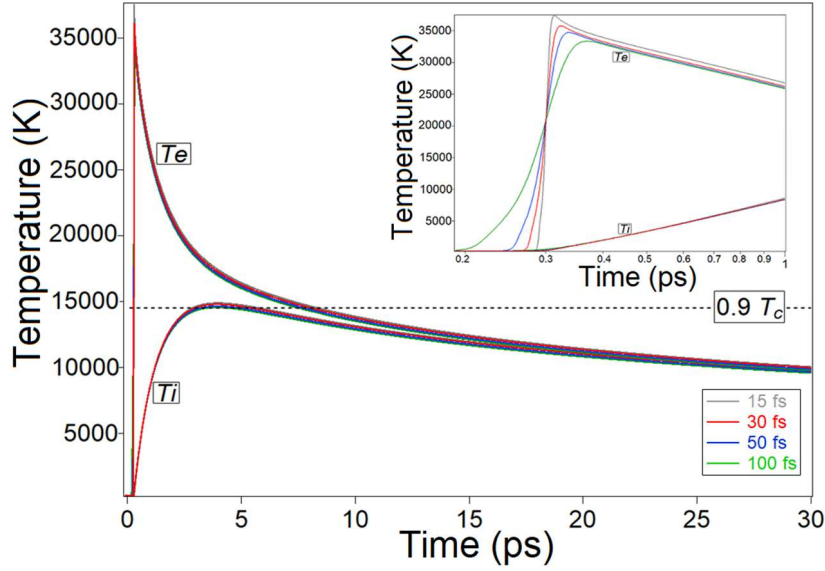

**Fig. 8** Evolution of electronic ($T_e$) and lattice ($T_i$) temperatures at the ablation threshold fluence ($F_{th,W}$ = 0.53 J/cm²) and pulse durations ranging from 15 to 100 fs for tungsten. The pulses are centered at 0.3 ps. The inserted frame corresponds to the evolution during the first picosecond.

The temperature reached by the lattice, regardless of the pulse duration, is $T_i \sim$ 14900 K. This value is higher than the melting (3695 K) and boiling temperatures (5828 K). Moreover, the critical temperature of tungsten being 16000 (+/-1000) K [41], the temperature criterion for phase explosion ($T_i \sim 0.9\ T_c$) is satisfied. The ablation threshold fluence measured for W may therefore correspond to phase explosion. However, photomechanical ablation cannot be excluded either. Our model (TTM) based only on temperature values does not distinguish the relative importance of the mechanisms. More sophisticated simulations using for example molecular dynamics [38] would be required.

As for Al, the values of $T_e$ and $T_i$ are further employed to compute the evolution of the electronic collision frequency (eqs. 3 and 4) and the evolution of reflectivity using the Drude-Lorentz model (eq. 2). The evolution of the reflectivity (and corresponding collision frequency) of a tungsten sample irradiated by a laser pulse at the fluence threshold is plotted on figure 9 for the four pulse durations. During the pulse, an increase of reflectivity from the measured unperturbed value (0.505) to a maximum value of ~ 0.535 is calculated. Then, for time above 0.4 ps (so after the end of the pulse), the electronic collision frequency continues to rise following the increase of the lattice temperature (fig. 8). This leads to a slow decrease of the reflectivity (fig. 9). To compare the simulation with our experimental results, the integral of the reflectivity during the pulse is calculated. For the four pulse durations, an integrated reflectivity value of $R \sim 0.52$ is obtained, which is again in good agreement with the experimental result also considering the measurement uncertainties (fig. 2).

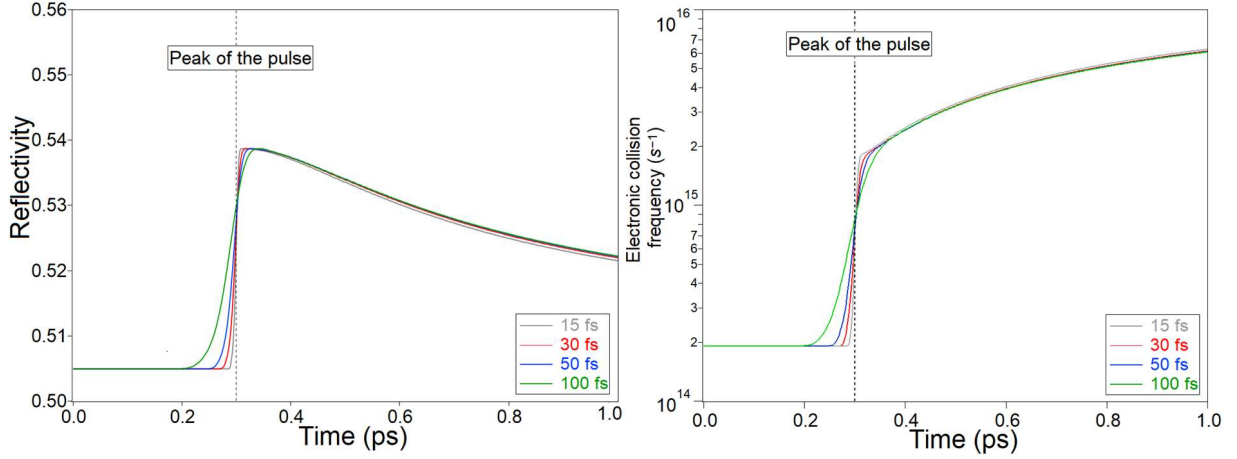

**Fig. 9** (a) Time evolution of the reflectivity of tungsten irradiated at the laser ablation threshold fluence ($F_{th,W}$ = 0.53 J/cm²). (b) Corresponding time evolution of the electronic collision frequency. The pulses are centered at 0.3 ps. Calculation done with the Drude-Lorentz model, based on the values of $T_e$ and $T_i$ obtained with the TTM.

### 5.3 Importance of DOS distribution for laser beam coupling

We now comment and summarize the main findings about the role of the DOS distribution in laser heating of Al and W in ultrashort laser interaction regime. For aluminum with a density of state close to the free electron gas model, the two-temperature model combined with the Drude model reproduces accurately the experimental results. Despite a relatively low number of available states for initial absorption of the laser photons (for $E \leq E_F$) which explains why the reflectivity keeps unchanged even above the ablation threshold, both laser coupling and electron heating are facilitated because of the availability of delocalized vacant states (s/p bands) above the Fermi energy. In those conditions, the chemical potential increases swiftly and the thermalization of the electronic population is rapid justifying the assumption made to apply the TTM model and to further calculate the reflectivity. Note that those conclusions related to the pertinence of the Drude and free electron gas models for describing the optical response of Al can be extended to higher fluences just following the scaling up of the effective collision frequency with excitation (as suggested from figs. 2 and 3 right).

Unlike aluminum, when the electronic collision frequency of tungsten increases, the reflectivity of the sample increases, as it is experimentally measured above the fluence threshold on figure 2. This behavior is due to the peculiar DOS of tungsten and its evolution following the excitation of the electronic population. A schematic diagram has been designed (fig. 10) to illustrate and summarize the laser heating of the electron population at various levels of excitation. Initially, the density of states close to the Fermi level and accessible to laser coupling is relatively low (fig. 10 (1)). When the pulse fluence is lower or equal to the ablation threshold, electronic transitions can take place in two possible ways only. Firstly, they can occur from the states located in the energy range: -1.20 eV < $E - E_F$ < 0 eV to the block of available states located beyond $E - E_F$ = 0.75. These transitions originating from a low DOS region (< 0.4 states/eV/atom) but facilitated by a large number of vacant states at higher energies are symbolized by a mid-thick green arrow (fig. 10 (2)). Secondly, the transitions can take place from the block located at $E - E_F$ < -1.2, symbolized by an orange arrow (fig. 10 (2)). In that case, because the reservoir of initial states is large but the density of arrival states is low, the transition process is therefore weak or very inefficient. In those conditions, the electronic population does not absorb enough energy to suppress the pseudo band-gap (but nonetheless, sufficiently at threshold to yield ablation) and the reflectivity keeps unchanged with respect to its unperturbed value (with $R = R_0 \cong 0.505$ and also $n_W = 3.63 + i2.81$).

For fluences above the ablation threshold, photon coupling becomes less effective. Indeed, the excitation of the electrons located below the pseudo band-gap is limited by the lack of available arrival states (thin orange arrows in figure 10 (3)). Meanwhile, inter-band transitions from the pseudo band-gap becomes more difficult because of its progressive filling (thin green arrow in figure 10 (3)). So, even if the energy phase space for the electron transitions is increasing (with the broadening of possible electron transitions towards high energy of the upper d-band), photon coupling is limited, and the reflectivity increases. This situation is in good agreement with the scenario depicted in [14, 42]. Experimentally, the maximum value reached by the reflectivity is: $R \sim 0.54$ (see figure 2) and almost the same value is obtained regardless of the pulse duration. Using the Drude-Lorentz model (equation 2) for convenient description of the permittivity in those irradiation conditions, this reflectivity corresponds to an optical index of $n_W = 4.475 + i2.99$ (in good agreement with [42]) and an effective collision frequency of $\nu_{eff} \sim 1.89 \times 10^{15}$ s$^{-1}$ whose increase is consistent with respect to the value determined in the unperturbed regime ($\nu_{eff} \sim 1.91 \times 10^{14}$ s$^{-1}$). This situation persists as long as the block of high-density of states above the pseudo band-gap is not sufficiently filled to provide an efficient source of coupling (in other words with many states available at the origin and arrival of the transition).

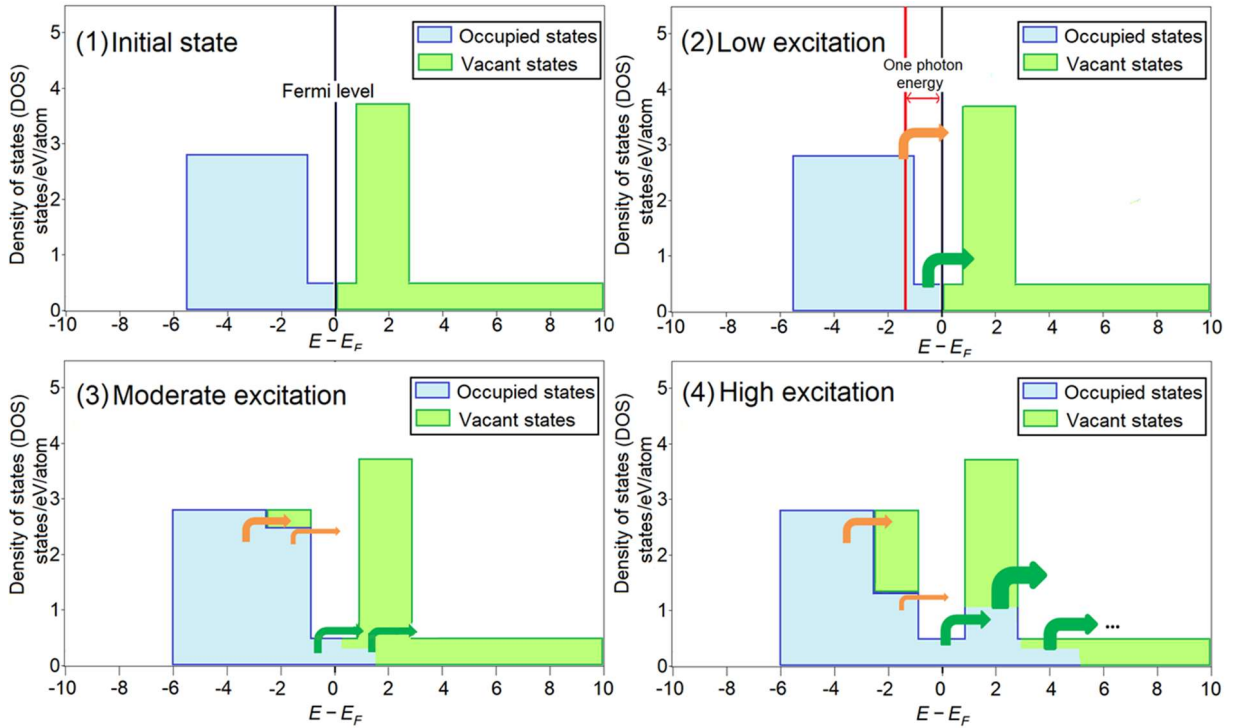

**Fig. 10** Schematic representation of tungsten DOS evolution at different excitation levels. The color and thickness of the arrows refer conceptually to the effectiveness of the transition: "high effective" transition in green, "low effective" transition in orange; the difference in width of the arrow is in accordance with the number of transitions that can occur.

For sufficiently high excitations, filling of the states above the Fermi level is more effective. Laser excitation can be significantly performed from these states (thick green arrows, figure 10 (4)) towards the higher energy states which density increases with the electronic temperature. The DOS distribution becomes particularly suitable for laser coupling with a large proportion of states available to absorb photons of the laser source. In these conditions, absorption becomes important and the reflectivity is reduced due to electron heating. As a free electron population is formed early during the pulse and s-band free electron transitions become rapidly dominant (the contribution of the d-band progressively vanishing, especially the one coming from its lower energetic part), one should apply without restriction the Drude model and the free electron gas approximation to describe tungsten at high excitation. For instance, for a measured reflectivity coefficient of $R \sim 0.41$ (*e.g.* reached for F $\cong$ 7.5 J/cm², 30 fs, see figure 2), only the Drude model is able to reproduce this experimental result ($R = 0.41$, with $v_{eff} \cong 5.03 \times 10^{15}$ s$^{-1}$ and corresponding complex index: $n_W = 1.47 + i1.97$, again in good agreement with the evolution shown in [42]). Note that performing the same calculation with a higher number of free electrons per atom consistently with the increase of accessible s/p transitions at high fluences [13] yields very close results.

The phenomenology here retraced on a wide range of excitation is in very good agreement with the experimental evolution of the reflectivity shown in figure 2. The interaction between tungsten and an ultrashort laser pulse can therefore be described using the two-temperature model and a hybrid permittivity modeling, using the Drude-Lorentz model at low and moderate fluences and further the Drude model at high fluences. Such approach provides a convincing evolution of the effective electron collision frequency and of the complex index and is able to correctly account for the peculiarity of the DOS distribution of a metal like tungsten. Finally, it is in very good agreement and also nicely complements former (mainly theoretical and simulation) works [13, 14, 42].

# 6 Conclusion

We measured the laser-induced ablation threshold and the evolution of the reflectivity of two metals (aluminum and tungsten) at 800 nm, for pulse durations of 15, 30, 50 and 100 fs. We first extended the ablation threshold scaling metrics to a range of duration poorly explored experimentally, demonstrating that the ablation threshold fluence is kept constant for both metals even to the range of few-optical-cycle laser pulses. It can be remarked that it corresponds to an intensity above $10^{13}$ W/cm² at 15 fs, becoming closer to what is measured for low band-gap dielectric material [43].

Another important observation is that the reflectivity measured at the ablation threshold fluence is equal to the unperturbed reflectivity for both metals despite the high intensity reached in our experiments performed in ultrashort regime (15 – 100 fs). To ascertain this outcome, it was verified numerically using the TTM and Drude-Lorentz model that the transient changes related to the different electronic temperature reached for all the pulse durations studied were not significant enough to alter the energy transfer to the lattice or the diffusive losses. Thus, as the amount of energy absorbed at the fluence threshold is the same for all four pulse durations, it justifies why a constant ablation threshold fluence was measured in our experiments.

As a feedback to modeling approaches used to describe laser heating and optical response, we conclude on the good applicability of the TTM model for both metals studied and of the Drude model for aluminum and Drude-Lorentz model for tungsten. In this framework, the approximation of an instantaneous thermalization of electron population reveals valid for both metals which DOS is relatively of low density around the Fermi level. Importantly, we determine that the particularity of DOS distribution of tungsten around the Fermi energy (d-band) must be taken into account using the Drude-Lorentz model to correctly retrace its optical response (reflectivity) in a wide range of excitation below and above the ablation threshold. Only at very high excitation ($\gg F_{th}$), when a free electron gas plasma is formed, the Drude model may apply in case of tungsten.

**Funding**

Financial support of the ASUR platform was provided by the European Community, Ministry of Research and High Education, Region Provence-Alpes-Côte d'Azur, Department of Bouches-du-Rhône, City of Marseille, CNRS, and Aix-Marseille University.

**Acknowledgments**

Thibault Genieys acknowledges the support of DGA – Direction Générale de l'Armement (Ministry of Defense) and Aix-Marseille University for his Ph'D grant.

All authors also thank Prof. E. Gamaly (Laser Physics Centre, Australian National University) and Dr. G. Tsibidis (IESL-FORTH) for valuable discussions on the topic of laser-matter interaction and CNRS-DERCI for their financial support through the International Research Project "IRP-MINOS".